\documentclass[aps,cha,twocolumn,groupedaddress]{revtex4}
\usepackage{graphicx}
\usepackage{amsmath}

\begin{document}
\title{Testing Dynamical System Variables for Reconstruction}
\author{T. L. Carroll}
\email{Thomas.Carroll@nrl.navy.mil}
\affiliation{US Naval Research Lab, Washington, DC 20375}

\date{\today}

\begin{abstract}
Analyzing data from dynamical systems often begins with creating a reconstruction of the trajectory based on one or more variables, but not all variables are suitable for reconstructing the trajectory. The concept of nonlinear observability has been investigated as a way to determine if a dynamical system can be reconstructed from one signal or a combination of signals  \cite{aguirre1995,letellier2005,aguirre2008,aguirre2011a,bianco2015}, however nonlinear observability can be difficult to calculate for a high dimensional system. In this work I compare the results from nonlinear observability to a continuity statistic that indicates the likelihood that there is a continuous function between two sets of multidimensional points- in this case two different reconstructions of the same attractor from different signals simultaneously measured. 

Without a metric against which to test the ability to reconstruct a system, the predictions of nonlinear observability and continuity are ambiguous. As a additional test how well different signals can predict the ability to reconstruct a dynamical system I use the fitting error from training a reservoir computer.
\end{abstract}
\pacs{ 05.45.-a, 05.45.Tp}

\maketitle

{\bf Analysis of a dynamical system often begins with reconstructing a trajectory for the system from one signal using a delay or differential embedding. In some cases, the signal picked for the reconstruction does not contain enough information about the entire system to make an accurate reconstruction. The concept of observability was initially developed for linear systems. Observability may be calculated from the Jacobian of a differential embedding based on one of the signals; if the Jacobian does not have the full rank of the dynamical system, the signal can not reproduce the full trajectory. The concept of observability was extended to nonlinear systems, but the nonlinearities can make calculation of the observability difficult for higher dimensional systems.

Continuity is a fundamental quantity from mathematics that can be used to determine if there is a continuous function $f$ between two sets of multidimensional data. Continuity can be used to answer the same question asked by observability.

Measures of continuity and observability for dynamical systems have been developed, but without a way to test whether these measures are correct, their application has been ambiguous. Reservoir computing is an offshoot of machine learning that trains a dynamical system to fit a signal based on an observation. For this work, a reservoir computer is used as an additional how accurately the observed signal can be used to reconstruct a dynamical system. The reservoir computer is driven with an input signal and is trained to fit one or more training signals. The fitting error is used as a measure of how well the training signals can be reproduced from the input signal. The three different types of measurement, observability, continuity of reservoir computers, measure different things, so their results do not always agree.
}

\section{introduction}
Can one reconstruct a nonlinear dynamical system based on observations of only one variable? Generically the answer from Takens' theorem is yes, but there are situations where a signal may not be able to reconstruct the dynamical system. Letellier, Aguirre and and collaborators have been developing various ways to use the concept of nonlinear observability to determine if a signal is sufficient for reconstructing an entire dynamical system \cite{aguirre1995,letellier2005,aguirre2008,aguirre2011a,bianco2015}. In some cases the nonlinear observability agrees with known properties of the dynamical system, but in other cases the interpretation of this statistic is more complicated. The difficulty of computing nonlinear observability for higher dimensional systems led to the development of symbolic observability \cite{bianco2015} or methods to compute observability from time series \cite{aguirre2008}.

In this paper I compare the symbolic observability to a continuity statistic that helps to indicate if there is a continuous function between two reconstructed dynamical systems \cite{pecora1995a}. It has been shown that reservoir computers can reconstruct dynamical systems \cite{lu2018}, so I use the fitting error of a reservoir computer \cite{manjunath2013} as an additional statistic against which to judge reconstructions. What I find is that the three different statistics can give different results because they measure different things. The observability statistic indicates if a differential embedding based on a particular variable is full rank, and continuity measures if one variable is predictable from another. There is as yet no good theory for reservoir computers, so is is hard to say what reservoir computers measure. 

By reconstructing a dynamical system from one signal, I mean that a signal $s(t)$ from a dynamical system has been digitized and stored. The dynamical system is reconstructed through the method of delays \cite{abarbanel1993}: a series of delay vectors is created from $s(t)$, where the vectors are ${{\bf{s}}_1} = \left[ {s\left( 1 \right),s\left( {1 + \tau } \right), \ldots s\left( {1 + \left( {d - 1} \right)\tau } \right)} \right],\;{{\bf{s}}_2} = \left[ {s\left( 2 \right),s\left( {2 + \tau } \right), \ldots s\left( {1 + \left( {d - 1} \right)\tau } \right)} \right], \ldots $ and so on, where the embedding dimension is $d$ and the embedding delay is $\tau$. The delay vectors, when plotted in a phase space, make up the reconstructed attractor.

First the three different statistics will be described. Nonlinear observability uses the equations for a dynamical system to determine if an embedding based on a particular variable has the same number of dimensions as the original dynamical system; if the embedding is lower dimensional, the full dynamical system can not be reconstructed. Continuity describes if knowing the location of a set of points on one dynamical system (which can be a full attractor or an embedding) leads to knowledge about where those points are on a different system, which could be the full system or an embedding based on a different variable. A reservoir computer is a high dimensional dynamical system, usually created by coupling together a network of nonlinear nodes. The reservoir computer is used to fit a training signal, and the statistic used is the error in fitting the training signal.

After the three statistics are described, they are applied to 5 different dynamical systems; a R{\" o}ssler system, a Lorenz system, a Chua system, a hyperchaotic R{\" o}ssler system and a H{\' e}non-Heiles system. The results are tabulated in two ways; first by comparing embeddings based on individual components to the entire dynamical system, and then by comparing embeddings based on one signal to embeddings based on a different signal from the same dynamical system.

\section{Nonlinear Observability}
Nonlinear observability is defined in \cite{hermann1977} (based on Kalman \cite{kalman1963}): 
 $\Sigma$ is a control system:
\begin{equation}
\label{control}
\begin{array}{l}
\quad \quad \dot x = f\left( {x,u} \right)\\
\Sigma: \\
\quad \quad s(t) = h\left( x \right)
\end{array}
\end{equation}
where $x$ represents the dynamical variables, $u$ is a control signal and $h(x)$ is an observer. If the dimension of the state space $m$ of $\Sigma$ is too small, then $\Sigma$ may not be able to distinguish between different states in the real system.

The observability can be calculated from the Lie derivatives of the observer.
\begin{equation}
\label{lie}
\dot s\left( t \right) = \frac{d}{{dt}}h\left( x \right) = \frac{{\partial h}}{{\partial x}}f\left( x \right) = {{{\cal L}}_f}h\left( x \right)
\end{equation}

${\cal L}_fh(x)$ is the Lie derivative of $h$ along the vector field $f$. The zero'th order Lie derivative is ${\cal L}_f^0h\left( x \right) = h\left( x \right)$ and the higher order Lie derivatives are
\begin{equation}
\label{lie_higher}
{\cal L}_f^jh\left( x \right) = \frac{{\partial {\cal L}_f^{j - 1}}}{{\partial x}}f\left( x \right)
\end{equation}
For a system with $m$ dimensions,
\begin{equation}
\label{obs_mat}
{O_s}\left( x \right) = \left[ {\begin{array}{*{20}{c}}
{\frac{{\partial {\cal L}_f^0h\left( x \right)}}{{\partial x}}}\\
 \vdots \\
{\frac{{\partial {\cal L}_f^{m - 1}h\left( x \right)}}{{\partial x}}}
\end{array}} \right]
\end{equation}
The system is observable if the rank of $O_s(x)=m$.

Letellier \cite{letellier2005} points out that the observability matrix $O_s(x)$ is also the Jacobian for a differential embedding of $x$. If this Jacobian matrix for a particular observer has singularities, then it will not be possible to reconstruct the system from that variable. Letellier uses an observability index $\theta_s(x)$ based on the eigenvalues of $O_s^TO_s$ and first defined in \cite{friedland1975},
\begin{equation}
\label{obs_index}
{\theta _s}\left( x \right) = \frac{{\left| {{\lambda _{\min }}\left[ {O_s^T{O_s},x\left( t \right)} \right]} \right|}}{{\left| {{\lambda _{\max }}\left[ {O_s^T{O_s},x\left( t \right)} \right]} \right|}}
\end{equation}
where the eigenvalues are evaluated at $x(t)$. The observability index for a differential embedding from a particular variable is the average of $\theta_s(x)$ over the entire trajectory.

Calculation of the observability for higher dimensional systems is complicated, so Bianco-Martinez et al. \cite{bianco2015} refined the concept of symbolic observability. Symbolic observability replaces the terms in the Jacobian of the differential embedding with symbols that divide the Jacobian terms into 4 types; null, constant, polynomial and rational. The observability can be calculated from the determinant of this symbolic Jacobian. The symbolic observability, denoted $\eta_s$, will be used in this paper.

\section{Continuity}
The definition of continuity is adapted from \cite{pecora1995a}. We have a mapping $f$ from a space $X$ to a space $Y$. We use $\left\| {} \right\|$ to indicate the Euclidean metric. The function $f$ is continuous at a point ${{\bf{x}}_0} \in X$ if for every $ \varepsilon  > 0$ there exists $\delta  > 0$ such that $\left\| {{\bf{x}} - {{\bf{x}}_0}} \right\| < \delta  \Rightarrow \left\| {f\left( {\bf{x}} \right) - f\left( {{{\bf{x}}_0}} \right)} \right\| < \varepsilon $. We proceed by choosing the $N_{\delta}$ nearest neighbors to $\bf{x}_0$. 

Our null hypothesis is that map ${\bf y}=f(\bf{x})$ from $X$ to $Y$ maps points randomly. We set the probability of a point within $\delta$ of $\bf{x}_0$ landing within a radius $\varepsilon$ of $\bf{y}_0$ as 0.5. We want to know the minimum radius $\varepsilon$ that is large enough to reject the null hypothesis. A total of $N_{\varepsilon}$ of the $N_{\delta}$ points will lie within a radius $\varepsilon$ of ${\bf y}_0$. To reject the null hypothesis with 95\% confidence, the  binomial distribution is used to find the minimum number of successes $N_{\varepsilon}$ in $N_{\delta}$ trials for which the area under the distribution is 0.95, if the probability of success on one trial is 0.5. As an example, if $N_{\delta}$ is 21, then $N_{\varepsilon}$ is 14. Table \ref{binomial} shows more examples of the binomial distribution.

\begin{table}[]
\centering
\caption{Example values of $N_{\delta}$ and $N_{\varepsilon}$ from a binomial distribution. $N_{\delta}$ is the number of points found within a radius $\delta$ of an index point on the attractor in the $X$ space, and $N_{\varepsilon}$ is the number of points on the attractor in the $Y$ space that are necessary to reject the null hypothesis that the points were mapped randomly from $X$ to $Y$.}
\label{binomial}
\begin{tabular}{|c|c|}
\hline
$N_{\delta} \quad$  & $N_{\varepsilon} \quad$     \\
\hline
 5 & 4 \\
 6 & 5 \\
 7 & 6 \\
 8 & 6 \\
 9 & 7 \\
 10 & 8 \\
 
\hline         
\end{tabular}
\end{table}

The size of $\varepsilon$ tells us how large of a radius the $N_{\delta}$ points must fall within to reject the null hypothesis (that the points were randomly chosen on the attractor) with a confidence of 95\%. If $\varepsilon$ is almost as large as the attractor, then there is probably not a continuous function between the embedded signal in the $X$ space and the embedded signal in the $Y$ space. A normalizing factor is necessary to get a value of $\varepsilon$ that is related to distance scales on the attractor. 

First, the set of $N_{\delta}$ points within the radius $\delta$ of $\bf{x}_0$ is found. The indices of these points are the set $I_{\delta}=(i_1, i_2 ... i_{N_{\delta}})$. Next, the set of distances ${D_y} = (\left\| {{\bf{y}}\left( {{i_1}} \right) - {{\bf{y}}_{\bf{0}}}} \right\|,\left\| {{\bf{y}}\left( {{i_2}} \right) - {{\bf{y}}_{\bf{0}}}} \right\|,...\left\| {{\bf{y}}\left( {{i_{{N_\delta }}}} \right) - {{\bf{y}}_{\bf{0}}}} \right\|)$ is found, where the $i$ indices are from the set $I_{\delta}$. From this set of distances, $\varepsilon$ is the distance to the $N_{\varepsilon}$'th  smallest distance in the set. 

The distance to the actual $N_{\varepsilon}$'th nearest neighbor to $\bf{y}_0$ as measured in the $Y$ space is $\varepsilon_0$. The normalized value of $\varepsilon$ is 
\begin{equation}
\label{eps_norm}
\varepsilon_n = \varepsilon_0 / \varepsilon .
\end{equation}
The normalizing factor $\varepsilon_0$ is used as the numerator so that smaller values of $\varepsilon_n$ indicate a lower probability of a function.The largest possible value of $\varepsilon_n$ is 1.0 This normalization is used to make it easier to compare the continuity statistic to the symbolic observability statistic.

The largest possible value of $\varepsilon_n$ is 1, which means that points that are nearest neighbors to $\bf{x}_0$ are also nearest neighbors to $\bf{y}_0$. A value of $\varepsilon_n$=1 indicates a high probability that $\bf{y}=f(\bf{x})$ is a continuous function. As $\varepsilon_n$ becomes smaller, the probability that $\bf{y}=f(\bf{x})$ is a continuous function is smaller. The mean value of $\varepsilon_n$ over the entire time series is
\begin{equation}
\label{mean_eps}
\Phi  = \frac{1}{N}\sum\limits_{i = 1}^N {{\varepsilon _n}\left( i \right)} .
\end{equation}

\begin{figure}
\centering
\includegraphics[scale=0.8]{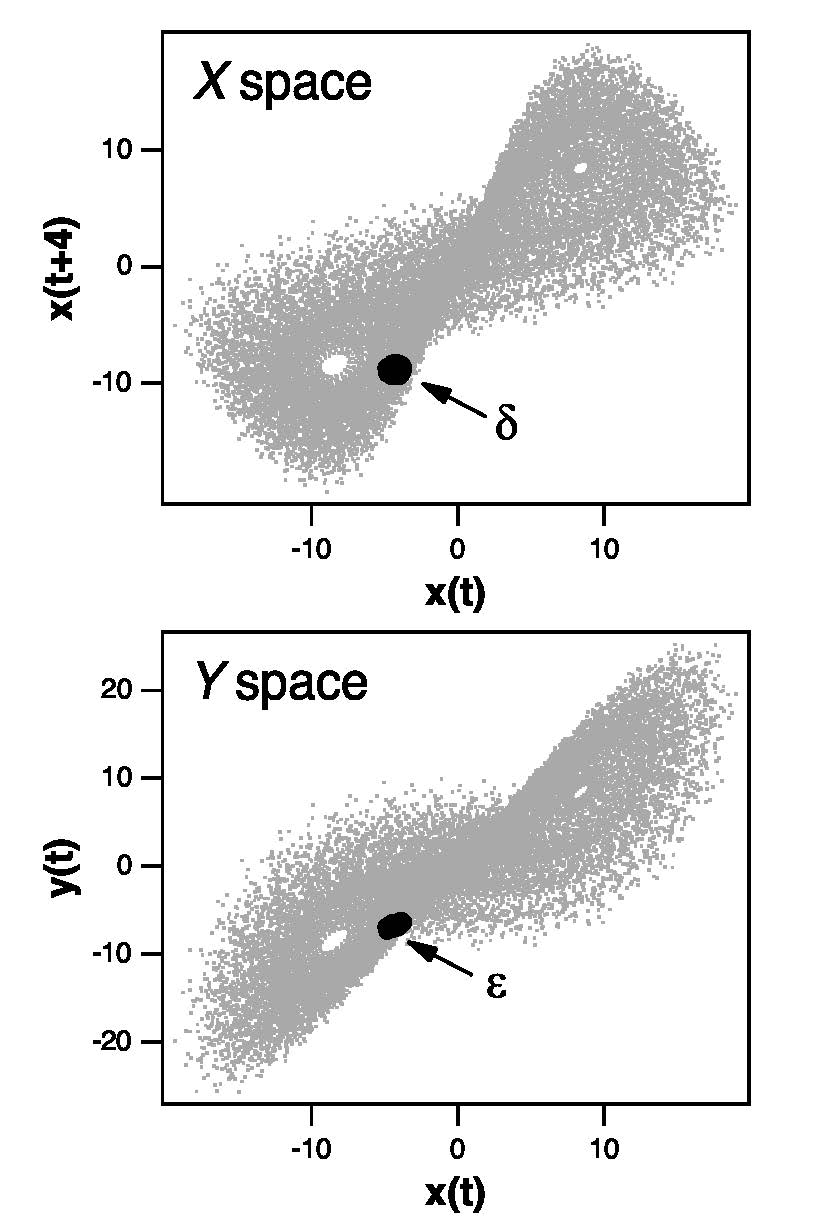}
\caption{\label{lorenz_del_eps} The top plot is the Lorenz attractor reconstructed from the $x$ variable with a delay of 4. The black points are a subset of the attractor that are within a radius $\delta$ of an index point. The bottom plot is the Lorenz $x$ and $y$ variables, so the bottom plot is a 2-d projection of the full Lorenz attractor. The black points on the bottom plot are the locations on the full attractor of the points within the $\delta$ radius on the top plot.}
\end{figure}

Figure \ref{lorenz_del_eps} shows an example of the radii $\delta$ and $\varepsilon$ for the Lorenz system. The top part of fig. \ref{lorenz_del_eps} shows the Lorenz attractor reconstructed from the $x$ signal with a delay of 4. The full attractor is shown in gray, while a subset of points that are within the radius $\delta$ of an index point is shown in black. The bottom part of fig. \ref{lorenz_del_eps} shows the the Lorenz $y$ signal plotted vs. the Lorenz $x$ signal, so the bottom plot is (a 2d- projection of) the full Lorenz attractor, not a reconstruction. The attractor is plotted in gray, while the black points show the location on the full attractor of the points that were within a radius $\delta$ of an index point on the attractor reconstructed from the $x$ signal. The set of black points on the bottom plot are used to find the radius $\varepsilon$.

\subsection{Sufficient data check}
 The continuity statistic does depend on having enough data to accurately represent the attractor. As a check on the validity of the continuity statistic, an over embedding statistic was also developed \cite{pecora2007}. The name "over embedding" is used because this statistic indicates that there is not sufficient data to embed the signal in $d$ dimensions.  A null hypothesis is proposed: the set of points within the $\delta$ radius of the index point were chosen randomly from all the points on the attractor. To test this null hypothesis, a histogram of inter-point distances $\Delta {\bf x}$ on the attractor is found. It is not necessary to include every inter-point distance; a large sub-sample of distances is enough. The histogram is normalized to create a probability distribution $\rho_{\delta}(\Delta {\bf x})$. 

Given a radius $\delta$, the probability that this $\delta$ value could have been obtained for a random set of points from the attractor is
\begin{equation}
\label{d_prob}
{p_\delta } = \int\limits_{x = 0}^\delta  {{\rho _\delta }\left( x \right)dx} .
\end{equation}

The probability that the null hypothesis is true, the points within $\delta$ were randomly distributed, is $p_{\delta}$. The confidence that the null hypothesis can be rejected is
\begin{equation}
\label{p_conf}
\Omega _{\delta }=1-p_{\delta}
\end{equation}

To reject the null hypothesis, we require $\Omega _{\delta } \ge 0.95$ .

The structure of an embedded signal may also affect the probability of getting a certain value of $\varepsilon$, so the analogous quantity for $\varepsilon$ is computed as $\Omega _{\varepsilon }$.

\subsection{Choosing neighborhood size}
In order to use the continuity statistic, it is necessary to choose some value for $N_{\delta}$, the number of neighbors in the $X$ space. The continuity statistic is a local measure- it measures the probability of a function between local regions on the attractor, so the radius $\delta$ should be small enough that it only encompasses local regions on the attractor. On the other hand, if $\delta$ is too small, it may be dominated by noise or digitization errors.

A clustering algorithm was developed in  \cite{carroll2016} that groups points by a statistic that may be loosely described as their "information content". I first pick a small group of neighboring points on the attractor. Does this group of points reveal anything about the structure of the attractor? If the group of points could have been sampled from a random distribution, then no information about the attractor is revealed, and I need include more neighboring points until I can eliminate the possibility that this set of points could have come from a random distribution.  Given a set of points, the algorithm in  \cite{carroll2016} finds how different in statistical terms that set of points is from a random distribution.

For a trajectory of length $N$, $N/10$ points are randomly chosen as index points, or centers for the small groups of neighbors. The radius $\delta$ is found around each index point by expanding a neighborhood about the index point and comparing the distribution of points in the neighborhood to the distribution that would be expected from a random distribution. A small region on the attractor is divided into $K$ equal size bins, and the number of points in each bin, $m_k$ is counted. The empirical probability of finding a point in each bin is 
${\hat \pi _k} = {{{m_k}} \mathord{\left/
 {\vphantom {{{m_k}} M}} \right.
 \kern-\nulldelimiterspace} M}$, where $M$ is the sum of the points in all $K$ bins. The model probability is a constant over all $K$ bins. Both sets of probabilities are used to update a prior containing the least information, and the posterior probabilities are compared using a  Kullback-Leibler divergence, \cite{kullback1951}, a commonly used measure of the difference between probability distributions. An analytic formula for this Kullback-Leibler divergence was derived in \cite{carroll2016}.  A penalty function of $K{\rm log_2}(K)$ must be subtracted from this divergence function, as creating more bins is the equivalent of overfitting the data. The final formula for measuring how different the posterior probability distribution inferred from the ${\hat \pi _k}$'s  from the posterior model distribution is

\begin{multline}
 \label{info}
R\left( {{m_k},K} \right) = \\
\frac{{\frac{1}{{\ln 2}}\sum\limits_{k = 1}^K {\left[ {({m_k} - {\rho _0}V) \cdot \psi ({m_k} + \frac{1}{2}) - \ln \Gamma ({m_k} + \frac{1}{2}) + \ln \Gamma ({\rho _0}V + \frac{1}{2})} \right]} }}{K}  \\
-\frac{{K{{\log }_2}\left( K \right)}}{K}
\end{multline}
 
 where ${\rho _0} = {{\sum\limits_{k = 1}^K {{m_k}} } \mathord{\left/
 {\vphantom {{\sum\limits_{k = 1}^K {{m_k}} } {\left( {KV} \right)}}} \right.
 \kern-\nulldelimiterspace} {\left( {KV} \right)}}$, where $V$ is the volume of an individual bin,  the function $\psi$ is the digamma function and $\Gamma$ is the gamma function. The units of $R(m_k,K)$ are bits/bin. A reasonable minimum threshold for $R(m_k,K)$ is 1 bit/bin. For this threshold, the attractor density is approximately constant over the $K$ bins.

Starting with one of the randomly chosen index points on the reconstructed dynamical system in the $X$ space, the $d+1$ nearest neighbors are located, where $d$ is the embedding dimension. Equation (\ref{info}) is used to find the value of $R\left( {{m_k},K} \right)$. If $R\left( {{m_k},K} \right) < 1$ bit, the neighborhood is expanded to include more points, and $R\left( {{m_k},K} \right)$ is calculated again. The expansion continues until $R\left( {{m_k},K} \right) \ge 1$ bit. The radius of this set of points is $\delta$ and the number of points in this set is $N_{\delta}$. From the set of $N_{\delta}$ points found on the embedded attractor in $X$, $N_{\varepsilon}$ of those points are nearest neighbors to the corresponding index point in $Y$, where $N_{\varepsilon}$ is determined from the binomial probability distribution. The continuity statistic may then be calculated from eq. (\ref{mean_eps}).

\section{Reservoir Computers}
It order to determine if the methods described above indicate when it is possible to reconstruct a dynamical system from a single variable, some type of reconstruction test is necessary. For this paper, different components of the dynamical system are reconstructed using a reservoir computer \cite{jaeger2004,lukosevicius2009,manjunath2013,van_der_sande2017,lu2018}. In \cite{lu2018}, it is shown that a reservoir computer can be used to reconstruct a dynamical system, suggesting that a reservoir computer is a way to test reconstruction algorithms.

Reservoir computing is a branch of machine learning. A reservoir computer consists of a set of nonlinear nodes connected in a network. The set of nodes is driven by an input signal, and the response of each node is recorded as a time series. A linear combination of the node response signals is then used to fit a training signal. Unlike other types of neural networks, the network connecting the nonlinear nodes does not vary; only the coefficients used to fit the training signal vary.

The reservoir computer used in this work is described by
\begin{equation}
\label{res_comp}
\frac{{d{\bf{R}}}}{{dt}} = \lambda \left[ {\alpha {\bf{R}} + \beta {{\bf{R}}^2} + \gamma {{\bf{R}}^3} + {\bf{AR}} + {\bf{W}}s\left( t \right)} \right].
\end{equation}
$\bf{R}$ is vector of node variables, ${\bf A}$ is a matrix indicating how the nodes are connected to each other, and ${\bf W}$ is a vector that described how the input signal $s(t)$ is coupled to each node. The constant $\lambda$ is a time constant, and there are $M=100$ nodes. For all the simulations described here, $\alpha=-3$, $\beta=1$ and $\gamma=-1$. The  matrix $\bf{A}$ is sparse, with 20 \% of its elements nonzero. The nonzero elements are chosen from a uniform random distribution between $\pm 1$, and then the entire matrix is normalized so that the largest real part of its eigenvalues is 0.5. Each row and each column of $\bf{A}$ has at least one nonzero element. The number of nodes used for these simulations was $M=100$.

The particular reservoir computer used here is arbitrary. The main requirements for a reservoir computer is that the nodes are nonlinear and that the network of nodes has a stable fixed point, so that in the absence of an input signal the network does not oscillate \cite{manjunath2013}. A different node type might yield different results, but the only way to determine this is by trial and error.

Equation (\ref{res_comp}) was numerically integrated using a 4'th order Runge-Kutta integration routine with a time step of 0.1.  Before driving the reservoir, the mean was subtracted from the input signal $s(t)$ and the input signal  was normalized to have a standard deviation of 1.

\begin{figure}
\centering
\includegraphics[scale=0.5]{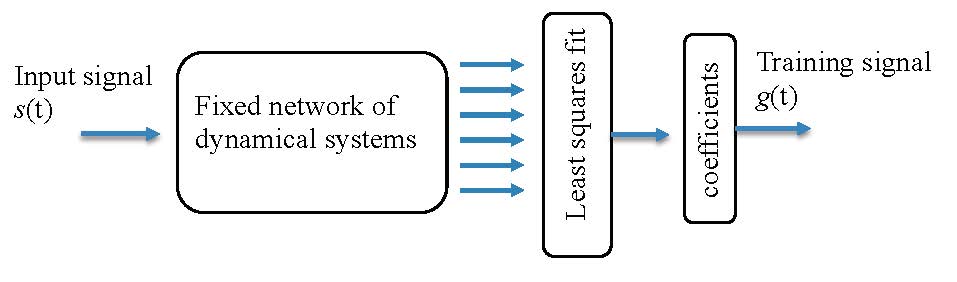} 
  \caption{ \label{reservoir_computer}
Block diagram of a reservoir computer. The input signal $s(t)$ drives a fixed network of dynamical nodes. The time varying signal from the nodes are fit to the training signal $g(t)$ by a least squares fit.}.
  \end{figure}

Figure \ref{reservoir_computer} is a block diagram of a reservoir computer. 

When the reservoir computer was driven with $s(t)$, the first 2000 time steps were discarded as a transient. The next $N=6000$ time steps from each node were combined in a $N \times (M+1)$ matrix
\begin{equation}
\label{fit_mat}
\Xi   = \left[ {\begin{array}{*{20}{c}}
{{r_1}\left( 1 \right)}&{{r_1}\left( 2 \right)}& \ldots &{{r_1}\left( N \right)}\\
 \vdots &{}&{}&{}\\
{{r_M}\left( 1 \right)}&{{r_M}\left( 2 \right)}& \ldots &{{r_M}\left( N \right)}\\
1&1& \ldots &1
\end{array}} \right]
\end{equation}
The last row of $\Xi $ was set to 1 to account for any constant offset in the fit. The training signal is fit by
\begin{equation}
\label{train_fit_0}
g\left( t \right) = \sum\limits_{j = 1}^M {{c_j}{r_j}\left( t \right)} 
\end{equation}
or

\begin{equation}
\label{train_fit}
{\bf{G}} = \Xi  {\bf{C}}
\end{equation}
where ${\bf{G}} = \left[ {g\left( 1 \right),g\left( 2 \right) \ldots g\left( N \right)} \right]$ is the training signal.

The matrix $\Xi $ is decomposed by a singular value decomposition
\begin{equation}
\label{svd}
\Xi   = {\bf{US}}{{\bf{V}}^T}.
\end{equation}
where ${\bf U}$ is $N \times (M+1)$, ${\bf S}$ is  $N \times (M+1)$ with non-negative real numbers on the diagonal and zeros elsewhere, and ${\bf V}$ is $(M+1) \times (M+1)$.

The pseudo-inverse of $\Xi $ is constructed as
\begin{equation}
\label{pinv}
{\Xi  _{inv}} = {\bf{V}}{{\bf{S}}^{\bf{'}}}{\bf{U}}
\end{equation}
where ${\bf{S}}^{\bf{'}}$ is an $(M+1) \times (M+1)$ diagonal matrix, where the diagonal element $S^{'}_{i,i}=S_{i,i}/(S_{i,i}^2+k^2)$, where $k=1 \times 10^{-5}$ is a small number used for ridge regression to prevent overfitting.

The fit coefficient vector is then found by
\begin{equation}
\label{fit_coeff}
{\bf{C}} = {\Xi  _{inv}}{\bf{G}}
\end{equation}.

The training error may be computed from
\begin{equation}
\label{train_err}
{\Delta _{RC}} = \frac{{\left\| {\Xi  {\bf{C}} - {\bf{G}}} \right\|}}{{\left\| {\bf{G}} \right\|}}
\end{equation}.
The training error is used as a measure of how well the training signal $\bf{G}$ may be reconstructed from the input signal ${s(t)} = \left[ {s\left( 1 \right),s\left( 2 \right), \ldots s\left( N \right)} \right]$.

The time constant $\lambda$ determined the frequency response of the reservoir. The time constant was adjusted to values between 0.1 and 6 to minimize the training error for different combinations of input and training signals.

\section{Comparison Between Different Statistics}
Symbolic observability and continuity statistics will be computed for several different chaotic systems to see if they can predict the training error from a reservoir computer. The observability statistic is based on differential or delay embeddings, while the continuity statistic is calculated for a delay embedding. Taking increasingly higher derivatives, necessary for a differential embedding, will lead to numerical problems for higher dimensional systems, which is why delay embeddings are used here.

There are two types of data tables presented below for the different dynamical systems. The first type of table for each system directly compares the symbolic observability $\eta_s$, the continuity $\Phi$ and the reservoir computer training error $\Delta_{RC}$. 

For the first type of table, we want to know how well the full dynamical system can be reconstructed from one of its individual variables. For the continuity statistic, this means we want to know how likely it is that there is a continuous function that maps an individual variable to the full system. The space $X$ is occupied by a delay embedding reconstructed from one of the individual variables, while the $Y$ space contains the full dynamical system. Larger values of the continuity statistic $\Phi$ indicate a greater likelihood that there is a continuous function between the delay reconstruction based on the individual signal and the full dynamical system. The maximum value of $\Phi$ is 1.

The symbolic observability statistic is also included in the first type of table because it indicates how well the full dynamical system can be reconstructed from one of its variables. Larger values of the symbolic observability $\eta_s$ indicate that there is a better chance the full dynamical system can be reconstructed from the individual variables. The maximum value of $\eta_s$ is 1. Finally, the first type of table contains the reservoir computer training error $\Delta_{RC}$ obtained by using one of the individual signals to drive the reservoir computer and fitting all the signals of the full dynamical system simultaneously.

The observability statistic determines how well the entire dynamical system may be reconstructed from a particular component. The reservoir computer and the continuity statistic, however, may also be used to indicate how well one component of a dynamical system may be recovered from a different component; for example, in \cite{lu2017}, a reservoir computer is used to fit individual components of the R{\"o}ssler or Lorenz systems.   For the second type of table, a delay reconstruction based on one of the variables from a dynamical system is compared to a delay reconstruction based on a different single variable from the dynamical system- not the full system, as in the first type of table. The second type of table reveals relationships between the individual components of the dynamical system. The second type of table shows the continuity statistic $\Phi$ computed for reconstructions based on individual variables from the dynamical system. The second type of table also shows the reservoir computer training error $\Delta_{RC}$ when the reservoir computer uses one signal from a dynamical signal as the input and fits a different individual signal from the same dynamical system, not multiple signals simultaneously, as in the first type of table. The second type of table also lists the confidence statistics for $\delta$ and $\varepsilon$, $\Omega_{\delta}$ and $\Omega_{\varepsilon}$. If either of the confidence statistics $\Omega_{\delta}$ or $\Omega_{\varepsilon}$ is less than 0.95, the continuity statistic $\Phi$ is not an accurate measure of the probability of a continuous function.

The collection of statistics is useful for determining if a particular component from a dynamical system is useful for reconstructing the full dynamical system, but in some cases neither the symbolic observability $\eta_s$ or the continuity $\Phi$ agree with the reservoir computer training error $\Delta_{RC}$. In these cases it is necessary to look at the actual signals themselves to see why the statistics may not be accurate. I will also speculate on why the reservoir computer training error $\Delta_{RC}$ does not always agree with the observability statistic.

\subsection{R{\"o}ssler System}
The R{\"o}ssler equations are \cite{rossler1976}
\begin{equation}
\label{rossler}
\begin{array}{l}
\frac{{dx}}{{dt}} =  - y - {p_1}z\\
\frac{{dy}}{{dt}} = x + {p_2}y\\
\frac{{dz}}{{dt}} = {p_3} + z\left( {x - {p_4}} \right)
\end{array}
\end{equation}

These equations were numerically integrated  with a time step $t_s$=0.1, and parameters $p_1=1$, $p_2=0.2$, $p_3=0.2$, $p_4=5.7$.

The symbolic observability indices for the R{\"o}ssler system are listed in table \ref{ross_full}. The observability matrix from the $y$ signal (eq. \ref{lie_higher}) is constant, so it has full rank for all values of $y$. It should therefore be possible to reconstruct the full state space of the R{\"o}ssler system from a measurement of the $y$ variable.

 The mean continuity statistic $\Phi$ (eq. \ref{mean_eps}) is also shown in table \ref{ross_cont}.   All the values of $\Phi$ for the R{\"o}ssler system are high, so there is a good probability of a continuous function, but the $z$ value is lower than for $x$ or $y$. 
 
 The continuity statistic $\Phi$ for the $x$ variable is larger than the continuity statistic for the $y$ variable, the opposite pattern of the observability  $\eta_s$. The reason is that the continuity statistic is not measuring the same thing as the observability. The continuity is a way of measuring predictability, which can be affected by the dynamics of the different signals as well as the rank of the embedding.
 
   It can be shown from the Jacobians for the differential embeddings that using the $x$ variable for a differential embedding expands volumes, while using the $y$ signal does not. The differential embedding Jacobian (the same Jacobian used to calculate observability) may be used to calculate exponents for the differential embedding in the same manner that Lyapunov exponents are calculated \cite{eckmann1985}; for a differential embedding based on the $x$ variable, these exponents are 11.1, 3.6 and 0.5 (in natural log units), while for the $y$ embedding the exponents are 1,0 and -1. The continuity statistic is measured by going from the embedding back to the full attractor, so if going from the full attractor to the $x$ embedding expands volumes, going in the reverse direction, from the $x$ embedding to the full attractor, contracts volumes. The $y$ embedding is neutral with respect to volume expansion or contraction. Because going from the $x$ embedding to the full attractor contracts volumes, the $\varepsilon$ radius on the full attractor is smaller when the $\delta$ radius is chosen on the $x$ embedding than when $\delta$ is on the $y$ embedding, so the continuity statistic appears larger for the $x$ signal than for the $y$ signal.

\begin{table}[]
\centering
\caption{Symbolic observability index $\eta_s$ from  \cite{bianco2015} , continuity statistic $\Phi$ from eq. (\ref{mean_eps}) and reservoir computer training error $\Delta_{RC}$ from eq. (\ref{train_err}), computed for the R{\"o}ssler system. For $\Phi$, the continuity was measured from an attractor reconstructed from the R{\"o}ssler signal in the $X$ column to the full  R{\"o}ssler attractor. For  $\Delta_{RC}$, the reservoir computer was driven by the signal in the $X$ column and all 3 signals from the R{\"o}ssler system were fit simultaneously}
\label{ross_full}
\begin{tabular}{|l|c|c|r|r|}
\hline
$X$ & $Y$   & $\eta_s$ \cite{bianco2015} & $\Phi$  & $\Delta_{RC}$  \\
\hline
   x      &  full system  & 0.88    & 0.26   &    $3.7 \times 10^{-4}$            \\
 
    y           & full system & 1.0    &  0.14  &   $4.2 \times 10^{-4}$       \\

z & full system &  0.44  & 0.09 & 0.048  \\

\hline         
\end{tabular}
\end{table}

Table \ref{ross_full} also shows the reservoir computer training error $\Delta_{RC}$ (eq. \ref{train_err}). The reservoir computer training error is much larger when the $z$ variable is used as an input to the reservoir computer than when the $x$ or $y$ variables are used as inputs, indicating that the $z$ variable does not work as well as the $x$ or $y$ variables for reconstructing the full R{\"o}ssler system.

Both the symbolic observability and the continuity statistic predict that the $z$ variable will be worse for reconstructing the state space of the full R{\"o}ssler system, and the training error from the reservoir computer confirms this. The ordering of the statistics is different, however; the symbolic observability statistic predicts that $y$ will be better for reconstruction than $x$, while the continuity statistic predicts that $x$ will be better than $y$. The reason for this discrepancy was explained above. The training error $\Delta_{RC}$ from the reservoir computer is almost the same when the $x$ or $y$ variable drives the reservoir.

\begin{table}[]
\centering
\caption{Continuity statistic $\Phi$ from eq. (\ref{mean_eps}) between a delay reconstruction from the variable in the $X$ space to a reconstruction from a variable in the $Y$ space, reservoir computer training error $\Delta_{RC}$ from eq. (\ref{train_err}), the confidence $\Omega _{\delta }$ (eq. \ref{p_conf}) that the radius $\delta$ in the $X$ space did not come from a randomly selected set of points in the attractor, and confidence $\Omega _{\varepsilon }$ that the radius $\varepsilon$ in the $Y$ space did not come from a randomly selected set of points in the attractor, computed for the R{\"o}ssler system. For $\Phi$, the continuity was measured from an attractor reconstructed from the R{\"o}ssler signal in the $X$ column to an attractor reconstructed from a signal in the $Y$ column. For  $\Delta_{RC}$, the reservoir computer was driven by the signal in the $X$ column, while the training signal was in the $Y$ column.}
\label{ross_cont}
\begin{tabular}{|l|c|c|r|r|r|}
\hline
$X$ & $Y$   &$\Phi$  & $\Delta_{RC}$ & $\Omega _{\delta }$ & $\Omega _{\varepsilon }$ \\
\hline
   x      & y      & 0.51   &    $3 \times 10^{-5}$   & 0.999  & 0.999    \\
   x             & z      & 0.18    &    $1 \times 10^{-3}$  & 0.999 & 0.59     \\
    y           & x      &  0.51  &   $6 \times 10^{-5}$   & 0.999  & 0.999   \\
y & z & 0.06 & $9 \times 10^{-4}$  & 0.999 & 0.59 \\
z & x & 0.41 & 0.058 & 0.85 & 0.999 \\
z & y & 0.19 & 0.062  & 0.85 & 0.999 \\
\hline         
\end{tabular}
\end{table}

\begin{figure}
\centering
\includegraphics[scale=0.8]{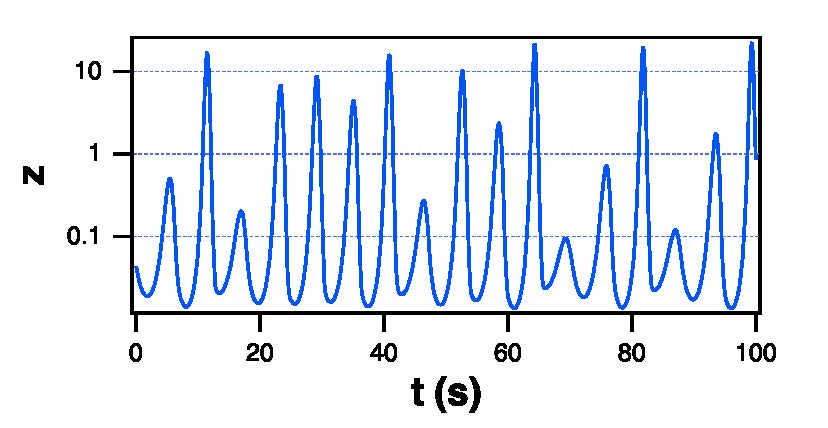} 
  \caption{ \label{ross_z}
Plot of the $z$ signal from the R{\"o}ssler system of eq. (\ref{rossler}). The signal is plotted on a logarithmic scale.}.
  \end{figure}

\begin{figure}
\centering
\includegraphics[scale=0.8]{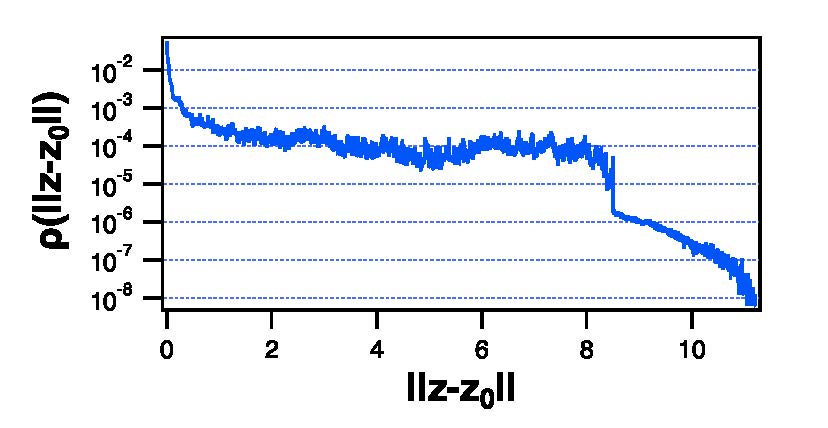} 
  \caption{ \label{ross_prob}
Probability $\rho \left( {\left\| {{\bf z} - {\bf z_0}} \right\|} \right)$ of the interpoint distances $\left\| {{\bf z} - {\bf z_0}} \right\|$ of the embedded $z$ signal from the R{\"o}ssler system of eq. (\ref{rossler}). }.
  \end{figure}

Table \ref{ross_cont} also shows that the reservoir computer training error $\Delta_{RC}$ is large for embeddings based on the $z$ variable even though the continuity statistic $\Phi$ is not small. The reason is that the confidence that the $\delta$ radius defined on the $z$ variable could not have resulted from a randomly selected set of points , $\Omega_{\delta}$, is only 0.85, which is below the threshold of 0.95 necessary for complete confidence that the continuity $\Phi$ is accurate. The confidence statistic $\Omega_{\delta}$ is small because of the structure of the $z$ variable.

. Figure \ref{ross_z} shows the $z$ signal from the R{\"o}ssler equations plotted on a logarithmic scale. Figure \ref{ross_prob} shows the probability $\rho \left( {\left\| {{\bf z} - { \bf z_0}} \right\|} \right)$ of the interpoint distances $\left\| {{\bf z} - {\bf z_0}} \right\|$. The probability distribution $\rho(\left\| {{\bf z} - {\bf z_0}} \right\|)$ has a maximum at small distances. When $N_{\delta}$ nearest neighbors are chosen on an embedding of the $z$ variable, there is a non-trivial probability that the distance $\delta$ could have been found from a set of points selected at random. Because the value of $\Omega _{\delta }$ is low (values $\le$ 0.95 are considered low), the continuity statistic is not reliable, so a high value of $\Phi$ does not establish that there should be a continuous function from the embedded $z$ variable to the embedded $x$ or $y$ variables.

The continuity statistic may also be calculated when the $X$ space is occupied by log($z$) and the Y space is occupied by the full attractor. In this case, the continuity from log($z$) to the full attractor is 0.75, indicating good continuity. The reservoir computer training error from log($z$) to the full attractor is $2 \times 10^{-4}$, an improvement over the $\Delta_{RC}$ = 0.048 in Table \ref{ross_full}.

Table \ref{ross_cont} also shows a low value for the continuity $\Phi$ from an embedded signal from the $y$ variable to an embedded signal from the $z$ variable, even though the reservoir computer training error $\Delta_{RC}$ for fitting $z$ from a reservoir computer driven by $y$ is small. Once again, the structure of the $z$ signal is responsible for the lack of reliability in the continuity statistic. The confidence $\Omega _{\varepsilon }$ that the value of the radius $\varepsilon$ in the $Y$ space could not have been found from a randomly selected set of points is only 0.59 whenever the $Y$ space contains $z$.

Again using log($z$) in the $X$ space instead of $z$, the continuity from log($z$) to $x$ is 0.75 and the continuity from log($z$) to $y$ is 0.46. When log($z$) is in the $X$ space, the confidence $\Omega_{delta}$ is 0.998, indicating a high level of confidence in the continuity statistic. The reservoir computer training error when the reservoir computer is driven by log($z$) and fits $x$ is $3.8 \times 10^{-4}$, while driving with log($z$) and fitting $y$ produces a training error of $1.8 \times 10^{-4}$. 

For the R{\"o}ssler system, the symbolic observability $\eta_s$ is the most useful statistic in predicting whether a particular variable may be used to reconstruct the full attractor. The continuity $\Phi$ is not always useful, but the over embedding statistics $\Omega _{\delta }$ and $\Omega _{\varepsilon }$ indicate when $\Phi$ is not useful.

\subsection{Lorenz}
The Lorenz equations are \cite{lorenz1963}
\begin{equation}
\label{lorenz}
\begin{array}{l}
\frac{{dx}}{{dt}} = {p_1}y - {p_1}x\\
\frac{{dy}}{{dt}} = x\left( {{p_2} - z} \right) - y\\
\frac{{dz}}{{dt}} = xy - {p_3}z
\end{array}
\end{equation}

with $p_1$=10, $p_2$=28, and $p_3$=8/3. The equations were numerically integrated with a time step of $t_s=0.02$.

The symbolic observability indices for the Lorenz system are listed in table \ref{lor_full}.

With an attractor reconstructed from a delay embedding occupying the $X$ space , the space $Y$ contained the full Lorenz attractor using all the variables from eq. (\ref{lorenz}). The embedding delay was 4. The mean continuity statistic $\Phi$ (eq. \ref{mean_eps}) is also listed in table \ref{lor_full}, as is the reservoir computer training error $\Delta_{RC}$. 

\begin{table}[]
\centering
\caption{Symbolic observability index $\eta_s$ from  \cite{bianco2015} , continuity statistic $\Phi$ from eq. (\ref{mean_eps}) and reservoir computer training error $\Delta_{RC}$ from eq. (\ref{train_err}), computed for the Lorenz system. For $\Phi$, the continuity was measured from an attractor reconstructed from the Lorenz signal in the $X$ column to the full  Lorenz attractor. For  $\Delta_{RC}$, the reservoir computer was driven by the signal in the $X$ column, all 3 signals from the Lorenz system were fit simultaneously}
\label{lor_full}
\begin{tabular}{|l|c|c|r|r|}
\hline
$X$ & $Y$   & $\eta_s$ & $\Phi$  & $\Delta_{RC}$  \\
\hline
   x      &  full system  & 0.78    & 0.37   &    $6.5 \times 10^{-4}$            \\
 
    y           & full system & 0.36    &  0.39  &   $3.6 \times 10^{-4}$       \\

z & full system &  0.36  &0.048 & 0.62  \\

\hline         
\end{tabular}
\end{table}

The symbolic observability index for $z$ is high, but the reservoir computer training error $\Delta_{RC}$ is also high. The symbolic observability index does not take into account the symmetry of the Lorenz equations; the equations are invariant under the transformation $\left( {x,y,z} \right) \to \left( { - x, - y,z} \right)$, so that the sign of $x$ and $y$ can not be determined from $z$. The continuity statistic $\Phi$ does detect this symmetry; the continuity statistic $\Phi$ for the $z$ signal to the full Lorenz system is only 0.048. The continuity statistic $\Phi$ is sufficient to determine which of the Lorenz variables is useful for reconstructing the full system.

\begin{table}[]
\centering
\caption{Continuity statistic $\Phi$ from eq. (\ref{mean_eps}) between a delay reconstruction from the variable in the $X$ space to a reconstruction from a variable in the $Y$ space, reservoir computer training error $\Delta_{RC}$ from eq. (\ref{train_err}), the confidence $\Omega _{\delta }$ (eq. \ref{p_conf}) that the radius $\delta$ in the $X$ space did not come from a randomly selected set of points in the attractor, and confidence $\Omega _{\varepsilon }$ that the radius $\varepsilon$ in the $Y$ space did not come from a randomly selected set of points in the attractor, computed for the Lorenz system.  For $\Phi$, the continuity was measured from an attractor reconstructed from the Lorenz signal in the $X$ column to an attractor reconstructed from a signal in the $Y$ column. For  $\Delta_{RC}$, the reservoir computer was driven by the signal in the $X$ column, while the training signal was in the $Y$ column.}
\label{lor_cont}
\begin{tabular}{|l|c|c|r|r|r|r|}
\hline
$X$ & $Y$   &$\Phi$  & $\Delta_{RC}$  & $\Omega _{\delta }$ & $\Omega _{\varepsilon }$ \\
\hline
   x      & y      & 0.79   &    $6.1 \times 10^{-4}$   & 0.999   & 0.999      \\
   x             & z      & 0.38    &    $1.2 \times 10^{-3}$    & 0.999  & 0.999   \\
    y           & x      &  0.42  &   $1.6 \times 10^{-4}$   & 0.999 & 0.999    \\
y & z & 0.17 & $6.8 \times 10^{-4}$ & 0.999  & 0.999 \\
z & x & 0.013 & 0.85 & 0.999 & 0.999 \\
z & y & 0.016 & 0.88 & 0.999 & 0.999 \\
\hline         
\end{tabular}
\end{table}

Table \ref{lor_cont} shows values of the continuity statistic $\Phi$ from eq. (\ref{mean_eps}) , reservoir computer training error $\Delta_{RC}$ from eq. (\ref{train_err}), and the over embedding statistics $\Omega _{\delta }$ and $\Omega _{\varepsilon }$ for single component embeddings of the Lorenz system. Table \ref{lor_cont} shows that larger values of the continuity statistic $\Phi$ usually correspond to smaller reservoir computer training errors $\Delta_{RC}$. The over embedding statistics $\Omega _{\delta }$ and $\Omega _{\varepsilon }$ are all well above 0.95, indicating that the continuity statistic $\Phi$ is dependable. The continuity from $y$ to $z$ is fairly low even though the reservoir computer training error $\Delta_{RC}$ is small.

\subsection{Chua System}
The Chua system is described by \cite{matsumoto1984}
\begin{equation}
\label{dscroll}
\begin{array}{l}
\frac{{dx}}{{dt}} = \alpha \left[ {y - x - f\left( x \right)} \right]\\
\frac{{dy}}{{dt}} = x - y + z\\
\frac{{dz}}{{dt}} =  - \beta y - \gamma z\\
f\left( x \right) = bx + 0.5\left( {a - b} \right)\left( {\left| {x + 1} \right| - \left| {x - 1} \right|} \right)
\end{array}
\end{equation}
with $\alpha=9$, $\beta=100/7$, $\gamma=0$,  $a=-8/7$ and $b=-5/7$. The integration time step was 0.05.

For calculation of the continuity statistic $\Phi$, the $X$ space contained an attractor reconstructed from a delay embedding of one of the components of the Chua system, with an embedding delay of 4. The $Y$ space contained the full attractor.  The results for the symbolic observability $\eta_s$, the continuity $\Phi$ and the reservoir computer training error $\Delta_{RC}$ are in table \ref{ds_full}.

\begin{table}[]
\centering
\caption{Symbolic  observability index $\eta_s$ from  \cite{bianco2015} , continuity statistic $\Phi$ from eq. (\ref{mean_eps}) and reservoir computer training error $\Delta_{RC}$ from eq. (\ref{train_err}), computed for the Chua system. For $\Phi$, the continuity was measured from an attractor reconstructed from the Chua signal in the $X$ column to the full  Chua attractor. For  $\Delta_{RC}$, the reservoir computer was driven by the signal in the $X$ column, all 3 signals from the Chua system were fit simultaneously}
\label{ds_full}
\begin{tabular}{|l|c|c|r|r|}
\hline
$X$ & $Y$   & $\eta_s$ & $\Phi$  & $\Delta_{RC}$  \\
\hline
   x      &  full system  & 0.78    & 0.27   &    $1.8 \times 10^{-3}$            \\
 
    y           & full system & 0.84    &  0.065  &   0.07       \\

z & full system &  1.0  &0.20 & $1.2 \times 10^{-3}$ \\

\hline         
\end{tabular}
\end{table}

The continuity statistic $\Phi$ and the training error $\Delta_{RC}$ produce similar results, but they disagree with the symbolic observability index. Both $\Phi$ and $\Delta_{RC}$ predict that the $x$ variable should give the best reconstruction of the Chua system, while $z$ should be less accurate and $y$ should give the worst reconstruction. The symbolic observability statistic, on the other hand, says that all 3 variables should give a good reconstruction.

\begin{table}[]
\centering
\caption{Continuity statistic $\Phi$ from eq. (\ref{mean_eps}) between a delay reconstruction from the variable in the $X$ space to a reconstruction from a variable in the $Y$ space, reservoir computer training error $\Delta_{RC}$ from eq. (\ref{train_err}), the confidence $\Omega _{\delta }$ (eq. \ref{p_conf}) that the radius $\delta$ in the $X$ space did not come from a randomly selected set of points in the attractor, and confidence $\Omega _{\varepsilon }$ that the radius $\varepsilon$ in the $Y$ space did not come from a randomly selected set of points in the attractor, computed for the Chua system. For $\Phi$, the continuity was measured from an attractor reconstructed from the Chua signal in the $X$ column to an attractor reconstructed from a signal in the $Y$ column. For  $\Delta_{RC}$, the reservoir computer was driven by the signal in the $X$ column, while the training signal was in the $Y$ column.}
\label{ds_cont}
\begin{tabular}{|l|c|c|r|r|r|}
\hline
$X$ & $Y$   &$\Phi$  & $\Delta_{RC}$ & $\Omega _{\delta }$ & $\Omega _{\varepsilon }$ \\
\hline
   x      & y      & 0.35   &    $3.5 \times 10^{-3}$   & 0.999  & 0.999       \\
   x             & z      & 0.48    &    $1.7 \times 10^{-3}$   & 0.999 & 0.999    \\
    y           & x      &  0.015  &   0.12   & 0.999   & 0.999  \\
y & z & 0.029 & 0.085 & 0.999 & 0.999 \\
z & x & 0.48 & $2.5 \times 10^{-3}$  & 0.999 & 0.999 \\
z & y & 0.43 & $8.1 \times 10^{-4}$ & 0.999 & 0.999 \\
\hline         
\end{tabular}
\end{table}

Table \ref{ds_cont} shows the continuity statistic and the training error for individual components of the Chua system. In table \ref{ds_cont}, both the continuity statistic $\Phi$ and the training error $\Delta_{RC}$ show that the $y$ variable is not good for reconstructing either the $x$ or $z$ components. 

\begin{figure}
\centering
\includegraphics[scale=0.5]{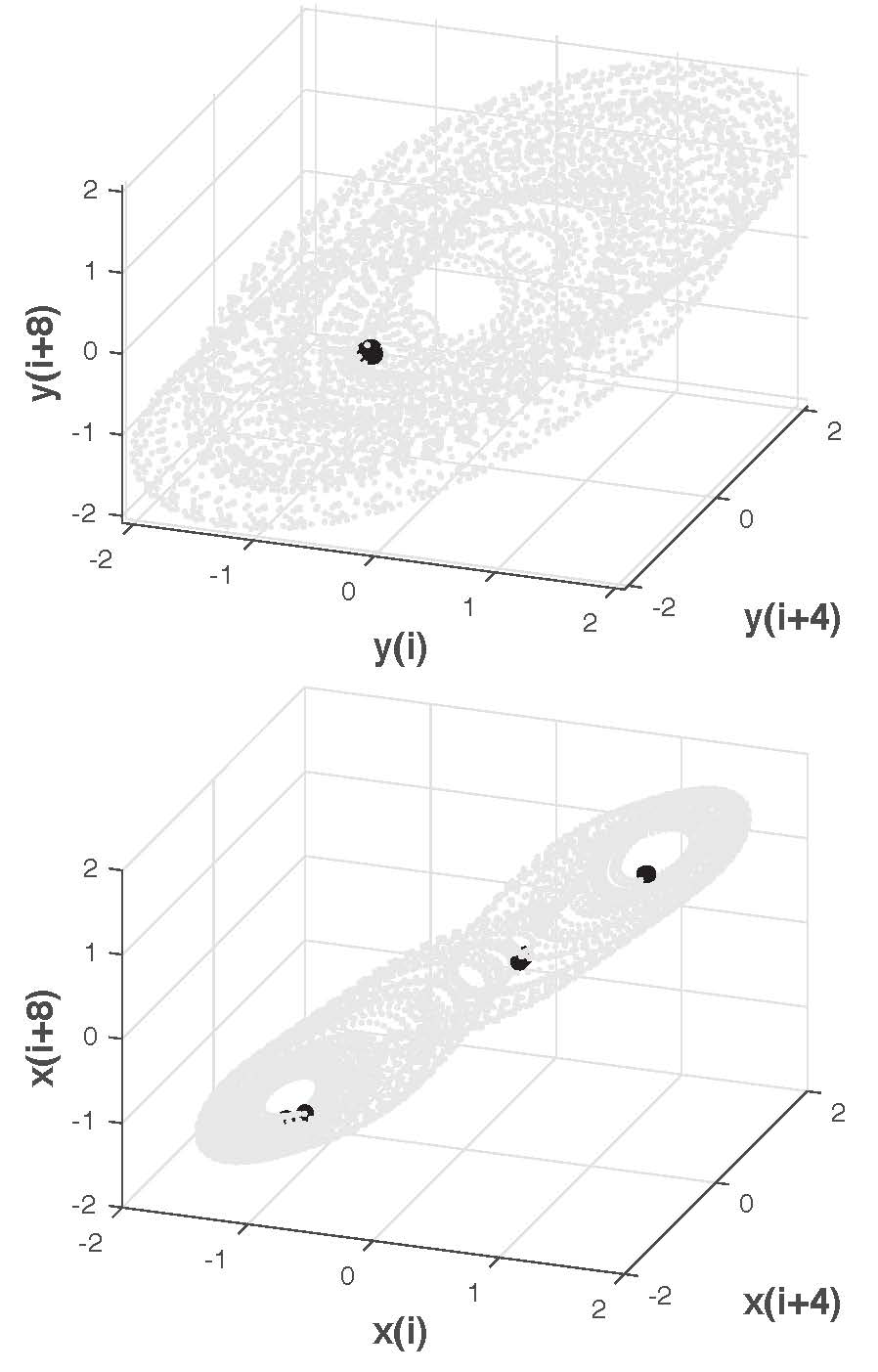} 
  \caption{ \label{double_scroll_y_and_x}
Top figure: The gray part of the figure is the Chua attractor reconstructed from an embedding of the $y$ signal. The black parts are a set of neighbors used to find a neighborhood of radius $\delta$ about an index point in order to calculate the continuity statistic $\Phi$. Bottom figure: The gray part of the plot is the Chua attractor reconstructed from an embedding of the $x$ signal, while the black points on the plot are the locations on the $x$ embedding of the points within a $\delta$ radius of an index point on the $y$ embedding. The radius of these points on the $x$ embedding determines $\varepsilon$ and the continuity statistic $\Phi$}.
  \end{figure}

Figure \ref{double_scroll_y_and_x}  shows why the $y$ variable from the Chua system is not good for reconstructing the Chua system. The top part of fig. \ref{double_scroll_y_and_x} shows a gray plot of the attractor created by embedding the $y$ variable, while the black points are the locations of the nearest neighbors used to find the radius $\delta$. The bottom plot in fig. \ref{double_scroll_y_and_x} shows an embedding based on the $x$ variable in gray. The points used in the $y$ embedding to find $\delta$ are shown in their corresponding positions on the $x$ embedding in black. Note that the points are located in both lobes of the attractor created from the $x$ signal, resulting in a large value of the radius $\varepsilon$ and therefore a small value of the continuity $\Phi$. This ambiguity in the Chua attractor is also why the reservoir computer training error $\Delta_{RC}$ is large when the $y$ variable drives the reservoir computer. The $y$ equation for the Chua system acts as a low pass filter on $x+z$. Filter inversion is known to be an ill-conditioned procedure, so it is not possible to recover $x$ or $z$ from the $y$ signal. The symbolic observability measures the rank of the embedding, which is not sensitive to this type of ambiguity.

\subsection{Hyperchaotic R{\"o}ssler System}
The various statistics may also be applied to higher dimensional systems. The hyperchaotic R{\"o}ssler system is described by \cite{rossler1979}
\begin{equation}
\label{hyper_ross}
\begin{array}{l}
\frac{{dx}}{{dt}} =  - y - z\\
\frac{{dy}}{{dt}} = x + ay + w\\
\frac{{dz}}{{dt}} = b + xz\\
\frac{{dw}}{{dt}} =  - cz + dw
\end{array}
\end{equation}
with $a=0.25$, $b=3$, $c=0.5$ and $d=0.05$. The equations were integrated numerically with a time step of 0.1.

\begin{table}[]
\centering
\caption{Symbolic observability index $\eta_s$ from  \cite{bianco2015} , continuity statistic $\Phi$ from eq. (\ref{mean_eps}) and reservoir computer training error $\Delta_{RC}$ from eq. (\ref{train_err}), computed for the hyperchaotic R{\"o}ssler system. For $\Phi$, the continuity was measured from an attractor reconstructed from the hyperchaotic R{\"o}ssler signal in the $X$ column to the full hyperchaotic R{\"o}ssler attractor. For  $\Delta_{RC}$, the reservoir computer was driven by the signal in the $X$ column, all 4 signals from the hyperchaotic R{\"o}ssler system were fit simultaneously}
\label{hcr_full}
\begin{tabular}{|l|c|c|r|r|}
\hline
$X$ & $Y$   & $\eta_s$ & $\Phi$  & $\Delta_{RC}$  \\
\hline
   x      &  full system  & 0.79    & 0.085   &    $9.2 \times 10^{ -3}$           \\
 
    y           & full system & 0.79     &  0.095  &   0.03       \\

z & full system & 0.44 & 0.06 & 0.26  \\

w & full system & 0.63 & 0.27 & 0.44  \\
\hline         
\end{tabular}
\end{table}

Table \ref{hcr_full} shows the symbolic observability index  $\eta_s$, the continuity statistic $\Phi$ and the reservoir computer training error $\Delta_{RC}$ when the $X$ space contained a delay embedding constructed from the signal in the $X$ column and the $Y$ space contained the full hyperchaotic R{\"o}ssler attractor.

\begin{table}[]
\centering
\caption{Continuity statistic $\Phi$ from eq. (\ref{mean_eps}) and reservoir computer training error $\Delta_{RC}$ from eq. (\ref{train_err}), computed for the hyperchaotic R{\"o}ssler system. For $\Phi$, the continuity was measured from an attractor reconstructed from the hyperchaotic R{\"o}ssler signal in the $X$ column to an attractor reconstructed from a signal in the $Y$ column. For  $\Delta_{RC}$, the reservoir computer was driven by the signal in the $X$ column, while the training signal was in the $Y$ column.}
\label{hcr_cont}
\begin{tabular}{|l|c|c|r|r|r|}
\hline
$X$ & $Y$   &$\Phi$  & $\Delta_{RC}$  & $\Omega _{\delta }$ & $\Omega _{\varepsilon }$\\
\hline
   x      & y      & 0.36   &    $3.8 \times 10^{-3}$  & 0.996  &   0.996      \\
   x             & z      & 0.34    &    0.016  &  0.996 & 0.38  \\
   x             & w      & 0.01    &    0.015  &0.996  & 0.974   \\
   
    y           & x      &  0.21  &   0.024   & 0.997   &  0.997 \\
y & z & 0.08 & 0.025 & 0.997  & 0.38 \\
y & w & $8 \times 10^{-3}$ & 0.054 &0.996 & 0.977 \\

z & x &  0.45 & 0.39 & 0.76 & 0.998 \\
z & y & 0.21 & 0.36 & 0.76 & 0.997 \\
z & w & $9 \times 10^{-3}$ & 0.035 & 0.76 & 0.993 \\

w & x & 0.064 & 0.34 & 0.986 & 0.996 \\
w & y & 0.039 & 0.32 & 0.986 & 0.996 \\
w & z & 0.064 & $6.7 \times 10^{-3}$ & 0.986 & 0.59\\

\hline         
\end{tabular}
\end{table}

Table \ref{hcr_cont} shows the continuity statistic and training error for different combinations of variables for the hyperchaotic R{\"o}ssler system. Both tables \ref{hcr_full} and \ref{hcr_cont} show that the reservoir computer training error is large when the $z$ or $w$ component is used as the input signal, even though the value of the continuity statistic $\Phi$ is large. The symbolic observability $\eta_s$ does predict that the $z$ variable is not good for reconstructing the full dynamical system, but the $w$ variable should be better, while table \ref{hcr_full} shows that the $w$ variable produces a larger training error when fitting the entire attractor.

\begin{figure}
\centering
\includegraphics[scale=0.8]{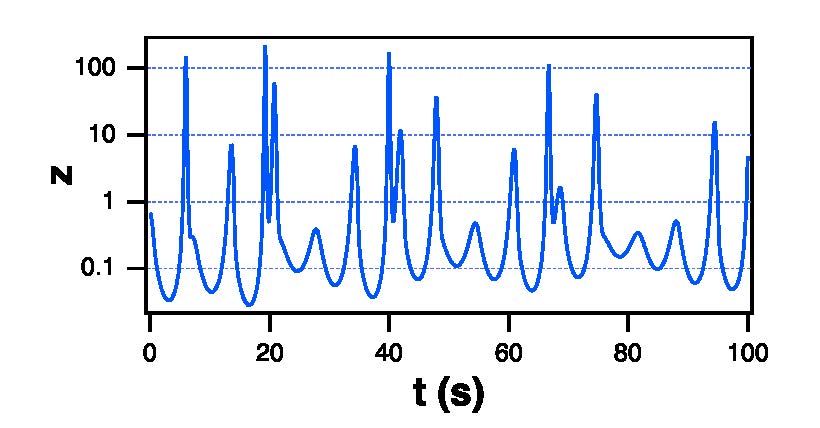} 
  \caption{ \label{hyper_ross_z}
Z signal from the hyperchaotic R{\"o}ssler system of eq. (\ref{hyper_ross}), plotted on a logarithmic scale.}.
  \end{figure}

Table \ref{hcr_cont} makes it clear that the $z$  variable produces large values of the training error $\Delta_{RC}$ even though the continuity statistic $\Phi$ is fairly large. Figure \ref{hyper_ross_z} shows that the hyperchaotic R{\"o}ssler $z$ signal resembles the regular R{\"o}ssler $z$ signal, in that it spends most of its time at small values with occasional large excursions. As a result,
the confidence that the group of points within a radius of $\delta$ in the $X$ space could not have been chosen randomly is rather low, 0.76. This low confidence means that there are not enough points to accurately sample the attractor constructed from an embedding of the $z$ variable, so the continuity statistic $\Phi$ when the $z$ variable is in the $X$ space is not accurate. The symbolic observability statistic also indicates that the $z$ variable is not useful for reconstructing the attractor. 

Table \ref{hcr_cont} also shows that while the continuity statistic $\Phi$ is small when the $X$ space contains the $w$ signal and the $Y$ space contains the $z$ variable, the reservoir computer training error $\Delta_{RC}$ is also small.  The confidence that the $\varepsilon$ radius could not have come from a randomly selected set of points is low for this comparison, only 0.59. Similarly, For the $y$ variable in the $X$ space and the $z$ variable in the $Y$ space, there is only a 38\% confidence that the radius $\varepsilon$ could not have come from a randomly selected set of points. This low confidence is again due to the particular structure of the $z$ signal.

\subsection{H{\'e}non-Heiles system}
The H{\'e}non-Heiles system was described by \cite{henon1964}
\begin{equation}
\label{hheiles}
\begin{array}{l}
\frac{{dx}}{{dt}} = u\\
\frac{{dy}}{{dt}} = v\\
\frac{{du}}{{dt}} =  - x - 2xy\\
\frac{{dv}}{{dt}} =  - y - {y^2} - {x^2}
\end{array}
\end{equation}
The H{\'e}non-Heiles system was conservative, so the initial condition was set to $x(0)=0$, $y(0)=0.67$, $u(0)=0.093$, and $v(0)=0$. The integration time step was 0.2.

Table \ref{hhr_full} does not show a strong correlation between either symbolic observability $\eta_s$ or continuity $\Phi$ and the reservoir computer training error $\Delta_{RC}$. Looking at individual components in table \ref{hcr_cont}, $\Delta_{RC}$ shows that $x$ and $u$ may be reconstructed from each other, or $y$ and $v$ may be reconstructed from each other, but trying to reconstruct other combinations of these variables does not work well.

\begin{table}[]
\centering
\caption{Symbolic observability index $\eta_s$ from  \cite{bianco2015} , continuity statistic $\Phi$ from eq. (\ref{mean_eps}) and reservoir computer training error $\Delta_{RC}$ from eq. (\ref{train_err}), computed for the H{\'e}non-Heiles system. For $\Phi$, the continuity was measured from an attractor reconstructed from the H{\'e}non-Heiles signal in the $X$ column to the full H{\'e}non-Heiles attractor. For  $\Delta_{RC}$, the reservoir computer was driven by the signal in the $X$ column, all 4 signals from the H{\'e}non-Heiles system were fit simultaneously}
\label{hhr_full}
\begin{tabular}{|l|c|c|r|r|}
\hline
$X$ & $Y$   & $\eta_s$ & $\Phi$  & $\Delta_{RC}$  \\
\hline
   x      &  full system  & 0.625    & 0.19   &    0.07           \\
 
    y           & full system & 0.625     &  0.13 &   0.34       \\

u & full system & 0.0 & 0.18 & 0.07  \\

v & full system & 0.0 & 0.19 & 0.34  \\
\hline         
\end{tabular}
\end{table}

\begin{table}[]
\centering
\caption{Continuity statistic $\Phi$ from eq. (\ref{mean_eps}) and reservoir computer training error $\Delta_{RC}$ from eq. (\ref{train_err}), computed for the H{\'e}non-Heiles system. For $\Phi$, the continuity was measured from an attractor reconstructed from the H{\'e}non-Heiles signal in the $X$ column to an attractor reconstructed from a signal in the $Y$ column. For  $\Delta_{RC}$, the reservoir computer was driven by the signal in the $X$ column, while the training signal was in the $Y$ column.}
\label{hh_cont}
\begin{tabular}{|l|c|c|r|r|r|}
\hline
$X$ & $Y$   &$\Phi$  & $\Delta_{RC}$  & $\Omega _{\delta }$ & $\Omega _{\varepsilon }$\\
\hline
   x      & y      & 0.21   &    0.19  & 0.998  &   0.998      \\
   x             & u      & 0.73    &   $7.3 \times 10^{-4}$  &  0.998 & 0.998  \\
   x             & v      & 0.20    &    0.015  &0.998  & 0.998   \\
   
    y           & x      &  0.1  &   0.51   & 0.998   &  0.998 \\
y & u & 0.11 & 0.51 & 0.998  & 0.998 \\
y & v & 0.68 & $2.1 \times 10^{-4}$ &0.998 & 0.998 \\

u & x &  0.68 & 0.012 & 0.998 & 0.997 \\
u & u & 0.24 & 0.17 & 0.998 & 0.997 \\
u & v & 0.25 & 0.19 & 0.998 & 0.997 \\

v & x & 0.21 & 0.49 & 0.998 & 0.997 \\
v & y & 0.67 & $2.7 \times 10^{-3} $ & 0.998 & 0.998 \\
v & u & 0.25 & 0.51 & 0.998 & 0.998\\

\hline         
\end{tabular}
\end{table}

\begin{figure}
\centering
\includegraphics[scale=0.4]{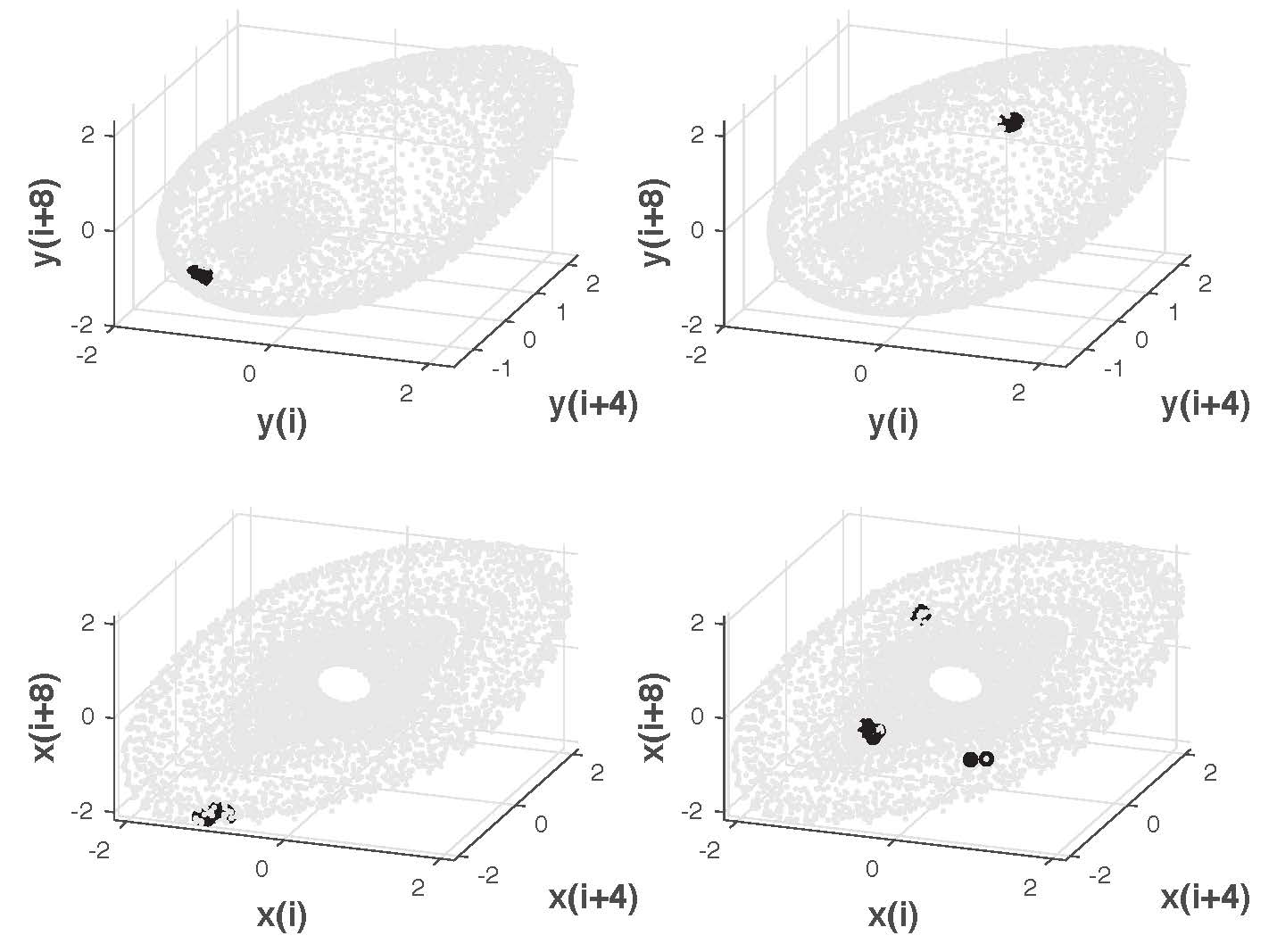} 
  \caption{ \label{henon_heiles_y_x}
Top left: attractor from the H{\'e}non-Heiles system $y$ variable embedded with a delay of 4. The black points are near neighbors located within a radius $\delta$ of an index point. Bottom left: attractor from the H{\'e}non-Heiles $x$ variable embedded with a delay of 4. The black points are the locations of the $x$ attractor of the black points from the $y$ attractor in the top left plot. Top right: same as the top left plot, but the points in black are chosen to be neighbors of a different index point. Bottom right: same as the bottom left plot, but the points in black are the locations on the $x$ attractor of the corresponding points on the top right $y$ attractor. The value of $\varepsilon$ for the bottom right plot will be much larger than the value of $\varepsilon$ for the bottom left plot.}.
  \end{figure}

Table \ref{hh_cont} also shows that for some combinations of variables, the continuity $\Phi$ can be large but the reservoir computer prediction error $\Delta_{RC}$ is also large. Figure \ref{henon_heiles_y_x} shows why there is a large error. On the left size, fig. \ref{henon_heiles_y_x} shows a plot of an embedding of the $y$ variable with points within the $\delta$ radius in black. On the left bottom, fig. \ref{henon_heiles_y_x} shows an embedding of the $x$ variable with the corresponding $\varepsilon$ neighborhood in black. The $\varepsilon$ neighborhood does appear to be split into 2 parts, but the 2 parts are close together, so $\varepsilon$ is not large. On the right side of fig. \ref{henon_heiles_y_x} are shown embeddings of the $y$ and $x$ variables again, but with the $\delta$ points chosen from a different neighborhood. In this case, the corresponding $\varepsilon$ points from the embedding of the $x$ variable (bottom right plot) are split into multiple regions, a clear indication that there is not a continuous function from $y$ to $x$. Figure \ref{henon_heiles_y_x} shows that sometimes the $\varepsilon$ region on the $x$ attractor is small, so the overall average value $\Phi$ is larger than would be expected. 

The H{\'e}non-Heiles system is conservative, which may be why large variations in continuity such as those seen in fig. \ref{henon_heiles_y_x} are seen.

\section{Conclusions}
\begin{figure}
 \includegraphics[scale=0.8]{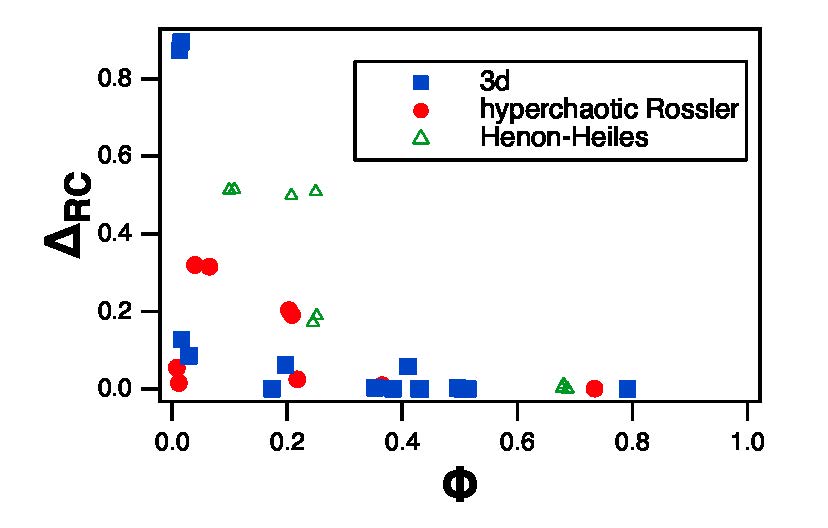} 
  \caption{ \label{err_vs_phi}
Reservoir computer training error $\Delta_{RC}$ vs. continuity $\Phi$. The 3d systems were the R{\"o}ssler, Lorenz and Chua systems. Comparisons where $\Omega _{\delta }$ or $\Omega _{\varepsilon }$ were $\le 0.95$ were excluded. }.
  \end{figure}

 To summarize the results for continuity, fig. \ref{err_vs_phi} shows the reservoir computer training error $\Delta_{RC}$ vs. the continuity $\Phi$, broken into 3 categories; the 3d systems (R{\"o}ssler, Lorenz and Chua), the hyperchaotic R{\"o}ssler system, and the H{\'e}non-Heiles system, excluding comparisons for which $\Omega _{\delta }$ or $\Omega _{\varepsilon }$ were $\le 0.95$. Except for the H{\'e}non-Heiles system, large values of $\Phi$ correspond to small values of $\Delta_{RC}$. If one had to set a threshold on $\Phi$ for getting an accurate reconstruction, $\Phi \ge 0.3$ would appear to be a reasonable value from fig. \ref{err_vs_phi}. As mentioned before, calculations of $\Phi$ may not be as accurate for the H{\'e}non-Heiles system because it is conservative.

\begin{figure}
 \includegraphics[scale=0.8]{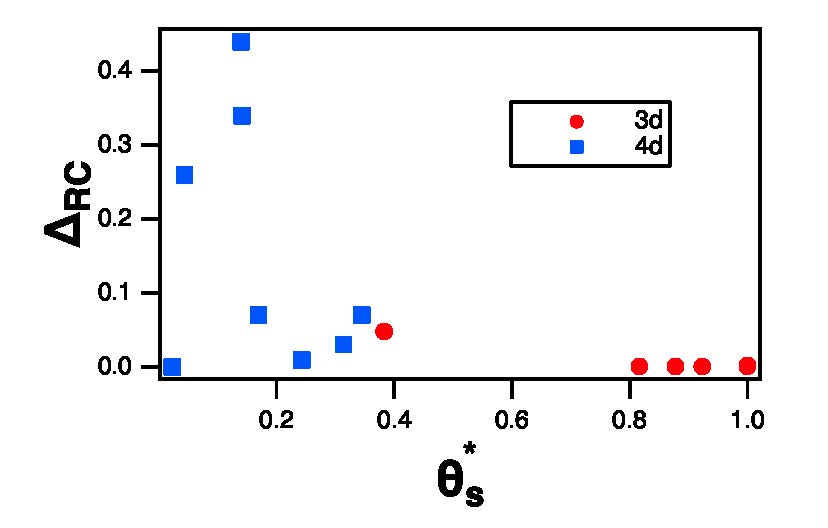} 
  \caption{ \label{obs_vs_err}
Reservoir computer training error $\Delta_{RC}$ vs. symbolic observability $\eta_s$. The 3d systems were the R{\"o}ssler, Lorenz and Chua systems, while the 4d systems were the hyperchaotic R{\"o}ssler and the H{\'e}non-Heiles systems. The Lorenz $z$ and Chua $y$ components were excluded because of symmetries that produced a large $\Delta_{RC}$. }.
  \end{figure}

  Figure \ref{obs_vs_err} summarizes the results for symbolic observability $\eta_s$. The Lorenz $z$ variable was excluded from the plot because the Lorenz equation is invariant under the transformation $(x, y, z)  \to (-x,-y, z)$, so the $z$ variable can not distinguish the sign of $x$ or $y$. For the Chua system, the $y$ variable is a low pass filtered version of $x+z$, so the contributions to the $y$ variable from the $x$ and $z$ variables can not be separated out, as shown in fig. \ref{double_scroll_y_and_x}. For the 3d systems, lower observability corresponded to larger reservoir computer training errors $\Delta_{RC}$. The situation is less straightforward for the 4d systems, but it has already been shown that the conservative nature of the H{\'e}non-Heiles system may make some of the statistics unreliable.
  
The three statistics described above measure different things. The observability statistic indicates whether a differential embedding based on a particular variable is full rank. The continuity statistic measures the ability to predict one signal based on knowing a different signal. Without a good theory, it is difficult to say what a reservoir computer measures.

Situations that lead to small values for the continuity statistic were explained above, and measures such as the overembedding statistics $\Omega _{\delta }$ or $\Omega _{\varepsilon }$ that indicate when the continuity statistic is not reliable were described. It is harder to explain why the observability statistic and the reservoir computer fitting error do not always agree. In some situations, such as the Lorenz $z$ variable, the differential embedding may be of full rank, but symmetries may make it impossible to reconstruct the full system. In other cases, the differential embedding may not be of full rank, but the reservoir computer training error is low. 

I may speculate on why the reservoir computer training error is low when the observability statistic indicates that a signal is less than full rank. As described in \cite{frunzete2012}, the embedding may not be of insufficient rank for all points on the attractor, but only for points on a singular manifold. It has been shown that reservoir computers can predict chaotic signals \cite{lu2018}; perhaps, if only a small subset of points lie on the singular manifold, the reservoir computer is able to fill in the gap; enough information is present that the reservoir computer can predict the missing information necessary to reconstruct the attractor.

Understanding the structure of the actual dynamical system is necessary to know when either of these statistics is not enough. If the equations for the dynamical system are not available, comparing embedded signals from different components can also reveal ambiguities, such as in fig. \ref{henon_heiles_y_x}.

\clearpage
\section{References}
\bibliography{measures_of_nonlinear_observability}{}

\end{document}